\documentstyle[12pt]{article}

\def\thefootnote{\fnsymbol{footnote}}
\def\Tr{{\rm Tr}}
\def\[{\left [}
\def\]{\right ]}
\def\({\left (}
\def\){\right )}

\def\pp{\partial}

\def\G{{\cal G}}

\newcommand{\be}{\begin{equation}}
\newcommand{\ee}{\end{equation}}
\newcommand{\bea}{\begin{eqnarray}}
\newcommand{\eea}{\end{eqnarray}}
\newcommand{\Lag}{{\cal L}}
\newcommand{\superint}{\int \diff^{4}\theta}
\newcommand{\lowest}{_{\theta =\bar{\theta}=0}}
\newcommand{\diff}{\mbox{d}}

\newcommand{\WaWa}{\Tr({\cal W}^{\alpha}{\cal W}_{\alpha})}

\newcommand{\DbDb}{{\cal D}_{\dot{\alpha}}{\cal D}^{\dot{\alpha}}}

\catcode`\@=11

\begin{document}
\begin{titlepage}
\begin{center}
      \hfill  LBNL-39744 \\
      \hfill  UCB-PTH-96/61 \\
      \hfill hep-th/9702105\\

{\large \bf Supersymmetry breaking and \\ 
weakly vs. strongly coupled string theory }\footnote{This work was 
supported in part by the Director, Office of
Energy Research, Office of High Energy and Nuclear Physics, Division
of High Energy Physics of the U.S. Department of Energy under
Contract DE-AC03-76SF00098 and in part by the National Science
Foundation under grant PHY-95-14797.}\\[.1in]

Pierre Bin\'{e}truy,\footnote{Visiting Miller Professor. Permanent address:
Laboratoire de Physique Th\'{e}orique et Hautes Energies (Laboratoire
associ\'e au CNRS--URA-D0063),
Universit\'{e} Paris-Sud, F-91405 Orsay, France}
Mary K. Gaillard\footnote{Miller Professor, Fall 1996.} and Yi-Yen Wu

{\em Department of Physics,University of California, and

 Theoretical Physics Group, 50A-5101, Lawrence Berkeley National Laboratory,
   Berkeley, CA 94720, USA}
\end{center}

\begin{abstract}
In the context of the field theory limit of superstrings, we consider an
almost realistic model of supersymmetry breaking by gaugino condensation
which includes, through nonperturbative corrections to the K\"ahler
potential, dilaton stabilization at a value compatible with a weak
coupling regime. Invariance under modular transformations is ensured
through a Green-Schwarz term and string threshold corrections, which
lead to moduli stabilization at the self-dual point. We are thus in a
position to discuss several issues of physical relevance: gravitino,
dilaton and moduli masses, axion, soft supersymmetry breaking parameters
and gauge coupling unification.
\end{abstract}
\end{titlepage}
\renewcommand{\thepage}{\roman{page}}
\setcounter{page}{2}
\mbox{ }

\vskip 1in

\begin{center}
{\bf Disclaimer}
\end{center}

\vskip .2in

\begin{scriptsize}
\begin{quotation}
This document was prepared as an account of work sponsored by the United
States Government. Neither the United States Government nor any agency
thereof, nor The Regents of the University of California, nor any of their
employees, makes any warranty, express or implied, or assumes any legal
liability or responsibility for the accuracy, completeness, or usefulness
of any information, apparatus, product, or process disclosed, or represents
that its use would not infringe privately owned rights. Reference herein
to any specific commercial products process, or service by its trade name,
trademark, manufacturer, or otherwise, does not necessarily constitute or
imply its endorsement, recommendation, or favoring by the United States
Government or any agency thereof, or The Regents of the University of
California. The views and opinions of authors expressed herein do not
necessarily state or reflect those of the United States Government or any
agency thereof of The Regents of the University of California and shall
not be used for advertising or product endorsement purposes.
\end{quotation}
\end{scriptsize}

\vskip 2in

\begin{center}
\begin{small}
{\it Lawrence Berkeley Laboratory is an equal opportunity employer.}
\end{small}
\end{center}

\newpage
\renewcommand{\thepage}{\arabic{page}}
\setcounter{page}{1}
\def\thefootnote{\arabic{footnote}}
\setcounter{footnote}{0}
%THIS IS PAGE 1 (INSERT TEXT OF REPORT HERE)
%\double

\section{Introduction}

One of the thorniest problems encountered by supersymmetry breaking mechanisms
in the context of superstring models is the stabilization of the dilaton field.
Indeed, as long as supersymmetry is conserved, the dilaton corresponds to a
flat direction of the scalar potential. Supersymmetry breaking lifts the
corresponding degeneracy and should therefore account for a realistic ground
state. This is a notoriously difficult problem. According to conventional
wisdom:
\begin{itemize}
\item the problem of supersymmetry breaking is fundamentally coupled to the
problem of dilaton stabilization, as was just explained.
\item only nonperturbative string effects can account for dilaton stabilization;
in other words only a precise knowledge of the strongly coupled string theory
will allow a solution to this problem. 
\end{itemize}
Once these two premises are accepted, it does not seem a far cry to conclude the
following:
\begin{itemize}
\item supersymmetry breaking can be understood only if one knows how to deal
with strongly coupled string theory.
\end{itemize}
Even in these days of duality and M-theory, this seems a remote possibility.
%And the curtain falls on act I with all principals being shot to death.
 
We argue in this article that, even though the former two statements are more
than plausible, one should not readily jump to the latter conclusion. Our
discussion will be based on an explicit model constructed in ref.~\cite{bgw2}
which agrees with the first two assertions but escapes the last one. This
model includes supersymmetry broken at a realistic scale, a stabilized dilaton,
moduli fields with couplings respecting modular invariance and a zero
cosmological constant. We believe that it is sufficiently realistic to allow
for a discussion of many issues associated with supersymmetry breaking and
moduli physics, based on actual computations rather than educated guesses.
Needless to say, we have no miraculous solution for either dilaton
stabilization or the vanishing of the cosmological constant. Although these
are incorporated in the model by fixing some parameters (only the second
constraint requires fine tuning), the model is still predictive enough in many
respects to provide a counter-example to the grim prospects mentioned above.

Before coming to the explicit model, let us explain why one may evade the
conclusion that supersymmetry is broken in a regime where string theory is
strongly coupled. Dilaton stabilization deals with the behavior of the scalar
potential at large and small values of the dilaton field. In what follows we
describe the dilaton as the lowest component $\ell$ of a linear supermultiplet
which also includes the antisymmetric tensor field: we believe that this is
the natural way to proceed,\footnote{This view is strongly reinforced by the
recent results of~\cite{deboer}.} at least when one deals with the weakly
coupled heterotic string. This is the formalism in which the Green-Schwarz
mechanism for anomaly cancellation is most easily implemented. It is also the
safest one when one starts to include nonperturbative effects --as we will--
since the equivalence between the linear and chiral multiplet formulations may
be blurred by these effects. 

Moreover $\ell$ is the string coupling: the string perturbation expansion can
be organized as a series in $\ell/(8 \pi^2)$. As we will see shortly, this
{\em string} coupling should not be confused with the coupling of the {\em
effective field theory}. In the limit of $\ell \rightarrow 0$, the string is
very weakly coupled and supersymmetry should be restored: one returns to the
flat direction and the scalar potential  vanishes\footnote{Presumably
$S$-duality does not help in this case. If it has anything to say, it is that
similarly in the limit of very strong coupling, the potential of the $S$-dual
theory vanishes.}. If one is prepared to cope with a supersymmetric minimum at
vanishing $\ell$, the real problem lies in the limit of strong coupling: $\ell
\rightarrow \infty$. It has been argued~\cite{bd1} that nonperturbative
corrections to the K\"ahler potential indeed stabilize the dilaton for large
$\ell$.  Our model is an  explicit  illustration of this
mechanism~\cite{bgw1,casas}.

This is important to get a stable minimum but it does not imply that this
minimum should lie in the large $\ell$ region. Indeed arguments based on the
unification of couplings tend to indicate that, if there is a physical ground
state, it lies in a region where at least the {\em effective theory} coupling
is small. Let us take this opportunity to stress that this is one of the few
predictions that one can infer from the low energy values of gauge couplings.
It is often stated that the precise value of the gauge coupling unification
scale ($3 \times 10^{16}$ GeV) is another one. We think that this is a
misleading statement since most models constructed so far that hold a claim
for being realistic include new forms of matter which perturb the evolution
of the gauge couplings at some intermediate threshold. 

In the model that we consider, nonperturbative contributions are
included in the K\"ahler potential in order to stabilize the dilaton.
They play an important role in providing a stable ground state. But
it is not necessary to choose unnatural values of the parameters in 
order to ensure that this ground state lies in the phenomenologically preferred
region of weak coupling. 

Let us briefly describe the explicit model that we use\footnote{ For more
details, we refer the reader to ref.~\cite{bgw2}.}. Supersymmetry is broken
through the condensation of gauginos associated with a hidden sector gauge
group $\G = \prod_a \G_a$, subgroup of $E_8$. We use the linear multiplet
formulation of gaugino condensation~\cite{bdqq,bgt} which introduces for each
gaugino condensate a vector superfield $V_a$. The dilaton field is the lowest
component of the vector superfield $V = \sum_a V_a$: $\ell = V|_{\theta = \bar
\theta = 0}$. The individual components $V_a|_{\theta = \bar \theta = 0}$ do
not appear in the effective theory component Lagrangian.

The gauge condensate superfields $U_a\simeq {\WaWa}_a$, where ${\cal W}_a$ is
the gauge chiral superfield for the group $\G_a$, appear as the chiral
projections of $V_a$:
\be
U_a\,=\,-(\DbDb-8R)V_a.
\ee
These condensates are taken to be static (nonpropagating) for reasons to be
discussed below.

The Lagrangian describing the gravitational sector including the dilaton 
simply reads:
\be
\Lag_{KE} = \superint \, E \left[ -2 + f(V) \right], \quad k(V) = \ln \, V + 
g(V), 
\ee
where $k(V)$ is the dilaton-dependent part of the K\"ahler potential and the
functions $f(V)$, $g(V)$ parameterize nonperturbative string effects. One might
wonder why one needs to introduce the function $f(V)$ besides the correction
$g(V)$. It turns out that the two are related by the condition
\be
V\frac{\diff g(V)}{\diff V}\,=\,
-V\frac{\diff f(V)}{\diff V}\,+\,f,
\label{eq:Einstein}
\ee 
which ensures that the Einstein term has canonical form.

Once these corrections are included, the {\em effective theory} coupling is
$g^2/2 = \ell / [1 + f(\ell)]$ (which is nothing but $(s + \bar s)^{-1}$ in the
dual chiral formulation). Thus if $f(\ell)$ is large, the string theory may be
strongly coupled when the effective field theory remains weakly coupled. It
turns out in our example that this is not the case at the ground state that we
consider (where we find $f(<\ell>)$ to be of order one).

The complete effective Lagrangian includes moduli fields and
allows for the presence of matter condensates described by chiral 
superfields $\Pi^\alpha$ (taken to be nonpropagating).
It includes the terms necessary to reproduce the modular 
anomaly~\cite{vy}-\cite{lt}
and two counterterms which allow the cancellation of this modular 
anomaly: the Green-Schwarz term~\cite{gs} and the term induced by 
string threshold corrections~\cite{dkl}\footnote{ However we do not include 
the moduli-dependent contribution from $N=2$ sectors that is not 
related to the Casimir operators of the gauge group~\cite{kkpr}.}. 
A superpotential for the matter condensates which respects the symmetries 
of the underlying theory is also added.

The K\"ahler potential for the effective theory is taken to be:
\be K = k(V) + \sum_I g^I,\;\;\;\;
g^I = -\ln(T^I + \bar{T}^I), 
\ee
and the complete effective Lagrangian reads: 
\bea 
\Lag_{eff} &=& \superint\,E \Bigg( -2 + f(V) + \sum_aV_a\Bigg\{
b'_a\ln(\bar{U}_aU_a/e^gV) \nonumber \\
& & + \; \; \; \sum_\alpha
b^\alpha_a\ln\(\Pi_r^\alpha\bar{\Pi}_r^\alpha\) 
- \sum_I{b_a^I\over8\pi^2}\ln\[\(T^I + \bar{T}^I\)
|\eta^2(T^I)|^2\]\Bigg\}\Bigg) \nonumber \\
 & & + \Bigg( {1\over2}\superint\,{E\over R}e^{K/2}W(\Pi^\alpha,T^I) +
{\rm h.c.} \Bigg)
\label{eq:Lag}
\eea
where $\Pi_r^\alpha = e^{\sum_I q^\alpha_I g^I/2} \Pi^\alpha$ is a 
modular invariant combination ($q^\alpha_I$ is the modular weight of
the matter condensate $\Pi^\alpha$). The coefficients $b'_a$ and
$b^\alpha_a$ are found by chiral and conformal anomaly matching.
The one-loop beta function coefficient is simply $b_a = b'_a +
\sum_{\alpha} b^{\alpha}_a$. One can extract the component Lagrangian
from (\ref{eq:Lag}) using standard procedures. Its bosonic terms
are given in ref.~\cite{bgw2} and we will not write them explicitly here.

Since the gaugino condensates $U_a$ are nonpropagating in this model, the
equations of motion fix their scalar components $u_a \equiv \rho_a e^{i
\omega_a}$ in terms of other fields. One might wonder whether some important
physics is missed by doing so. We argue that this is the only correct
procedure. The dynamical case has been studied in detail in ref.~\cite{yy}
using the simpler example of an $E_8$ gauge condensate. One may check for this
example that both the condensate value $\rho$ and its phase $\omega$ are
fields of mass larger than the condensate scale. In order to be consistent,
one should therefore integrate over them, in which case one recovers the
theory with a {\em static} $E_8$ condensate~\cite{bgw1}\footnote{ If we take
the example of QCD, the same should be correct for the matter condensates, at
least in the absence of Goldstone bosons associated with broken global chiral
symmetry.}. We believe that this is general and that the only consistent
effective theory below the condensation scale is the static theory. Hence the
only dilaton-like scalar and axion-like pseudoscalar in this effective theory
are the model independent dilaton and axion of the original string theory
(barring some small mixing with the heavy static condensates and some high
order mixing with the moduli).

One finds:
\be \rho^2_a = e^{-2{b_a'\over b_a}}e^Ke^{-{(1+f)\over b_a\ell} - 
{b\over b_a}\sum_Ig^I}\prod_I|\eta(t^I)|^{{4(b-b_a) \over b_a}}
\prod_\alpha|b^\alpha_a/4c_\alpha|^{-2{b_a^\alpha \over b_a}},
\label{eq:condensate} 
\ee
where $b = C_{E_{8}}/(8\pi^2)$, $C_{E_{8}}$ is the quadratic Casimir operator 
in the adjoint of $E_8$,
and $c_{\alpha}$ are couplings in the superpotential, presumably of order one.

This is an interesting formula in many respects. One recovers in particular a
behavior in $e^{-2/b_a g^2}$ once one identifies correctly the squared gauge
coupling of the effective theory as given by $2\ell/(1+ f(\ell))$.

This behavior reveals the limitations of the dual formulation using the chiral
supermultiplet $S$ to describe the string dilaton. In such a formulation, the
gaugino condensate is described by a chiral superfield $H^3$ which is found to
depend on the dilaton as $e^{-S/ b_a}$. This seems to be in agreement with the
preceding behavior since a duality transformation yields {\em at lowest
order}:
\be
{1 + f(L) \over L} = S + \bar S
\label{eq:duality}
\ee
and it is often attributed to the ``power of holomorphy''.

However powerful, holomorphy should be taken with a grain of salt in this
instance  because higher order corrections modify the relation
(\ref{eq:duality}). In particular, the Green-Schwarz term contributes to the
right-hand side in a nonholomorphic way. This is the standard observation that,
in the chiral formulation, the expansion parameter which is given in terms of
the dilaton field as $1/ {\rm Re} \, S$ needs to be redefined at each order of
perturbation theory.

One may wonder how the chiral superfield $H^3$ dependence in $S$ may undergo
these nonholomorphic redefinitions. The point is that it is incorrect to take
$H^3$ as an unconstrained superfield because it describes $\WaWa$ in the
effective theory, and this field is subject to the usual Bianchi identities of
the gauge sector. The correct treatment is through the use of a 3-form
supermultiplet~\cite{bggp}  and is naturally implemented in the linear
multiplet formulation~\cite{bgt}.

The scalar potential reads:
\be 16\ell^2 V = 
\( 1+\ell {dg \over d\ell} \)\left|\sum_a\(1+b_a\ell\)u_a\right|^2 -
3\ell^2\left|\sum_ab_au_a\right|^2
+ 4\ell^2\(1+b\ell\)\sum_I\left|{F^I\over{\rm Re}t^I}\right|^2. 
\label{eq:potential}
\ee
where 
\be
F^I = \sum_a {u_a(b - b_a) \over 4 (1+b\ell)^2} \(1 + {4 \over
\eta(t^I)} {\partial \eta \over \partial t}(t^I){\rm Re}t^I\)
\ee 
is the $T^I$ auxiliary field. 

The minimum with respect to the modulus $t^I$ is obtained 
at the self-dual point $t^I = 1$ where $F^I = 0$. This has some important 
phenomenological consequences as we will see below.

As for the remaining terms, the potential appears to be dominated by the
condensate with the largest one-loop $\beta$-function coefficient, so the
general case is qualitatively very similar to the single condensate
case\footnote{ This is quite different from racetrack models~\cite{racetrack}
where one has to play one condensate against another.}, and it appears that
positivity of the potential can always be imposed. One thus does not need to
appeal to another source of supersymmetry breaking to cancel the cosmological
constant. 

The gravitino mass is found to be 
\be
m_{\tilde G} = {1\over3}\langle|M|\rangle =
\frac{\, 1}{\, 4\,}\,\langle|\sum_ab_au_a|\rangle
\ee
where $M$ is the supergravity auxiliary field. The mass
is thus also governed by the {\em vev} of the condensate with the largest 
beta-function coefficient. The scale of supersymmetry breaking is therefore 
found to be naturally low as long as the gauge group is smaller than $E_8$.
There is in fact a further reduction due to the dependence of
the condensate (\ref{eq:condensate}) on the moduli: 
at the self-dual point the reduction factor is
approximately $e^{-\pi(b - b_a)/2b_a}$.

Of course, what plays an important role in stabilizing the dilaton
as well as fine-tuning the cosmological constant to zero is the exact form
of the functions $f(\ell)$ and $g(\ell)$. They are constrained by 
(\ref{eq:Einstein}) and subject to the weak coupling boundary conditions:
\be
f(\ell=0)=0,\; \; \; g(\ell=0)=0.
\ee
Thus $g(\ell)$ is completely determined once we have chosen $f(\ell)$.
Also, the requirement of boundedness of the potential (\ref{eq:potential}) in
the strong coupling limit gives:
\be
\( \ell {dg \over d \ell} \) (\ell \rightarrow \infty) \ge 2.
\label{eq:stab}
\ee
Two possible choices for the function $f$ are~\cite{bd1} $f = A e^{-B/\ell}$
and~\cite{shenk} $f=A_p (\sqrt{\ell})^{-p} e^{-B/\sqrt{\ell}}$. Since all the
constraints on the functions $f$ and $g$ are invariant under a rescaling of
$\ell$, the coefficient $B$ can be fixed to obtain the right value of $< \ell
>$ as measured by the gauge coupling at unification. And the parameter $A$ (or
$A_p$) can be used to fine-tune the cosmological constant to zero.
 
One may look more closely at the second case which is a genuine stringy
nonperturbative effect\footnote{ We do not consider here the case where the
coefficient $B$ in the exponent is  moduli-dependent~\cite{eva}. If such
nonperturbative stringy contributions turn out not to be modular invariant,
this would perturb the moduli ground state away from the self-dual point.}.
Taking for  illustrative purposes $f=[A_0 + A_1  \ell^{-1/2}]\,
e^{-B\ell^{-1/2}}$, where the condition (\ref{eq:stab}) requires $A_0$ to be
larger than $2$, one finds a realistic minimum for values of the parameters of
order one: $B<\ell>^{-1/2} \sim 1.1$ to $1.3$, $A_0 \sim 2.7$ to $5.3$ and $A_1
\sim -3.1$ to $-4.6$.

One particularly interesting aspect of the model is axion physics. Pseudoscalar
fields are the phases $\omega_a$ of the condensates and the so-called
model-independent axion which is dual to the fundamental antisymmetric tensor
field. The latter couples in a universal way to the 
$F^{a\mu \nu}{\tilde F}_{a \mu
\nu}$ term of each gauge group factor. If again we look at the dynamical model
with one $E_8$ condensate~\cite{yy} we find that out of the two possible
pseudoscalars, the condensate phase is very heavy whereas the model-independent
axion remains massless. This is obviously the supersymmetric counterpart of
what happens with the scalars. Again, it justifies our approach which treats the condensate degrees of freedom
as static in the effective theory. If we allow for more than one
condensate, the model-independent axion acquires a
very small mass (typically exponentially suppressed relative to the gravitino
mass by a factor of order $\sqrt{\rho_2 / \rho_1}$ in the two-condensate
case~\cite{bgw2}). Thus we are left with a very light pseudoscalar which has
the right couplings to be the QCD axion. This was actually noted by
Banks and Dine~\cite{bd1} who used an argument based on the breaking of a
continuous $R$-symmetry. Our model provides an explicit realization of this
phenomenon.  Because of corrections to its gauge kinetic term, the
model-independent axion must be normalized properly, which gives a reduction
factor for the axion decay constant $f_a$ equal to $\ell \sqrt{2[1 + \ell
dg/d\ell]}$ in reduced Planck mass units ($m_{Pl}=1$). This factor is
approximately equal to $b_a \ell^2 \sqrt{6}$ at the vacuum for the single
gauge condensate case studied in~\cite{bgw2}. This gives a suppression factor of
about $1/50$ if the gravitino mass is found around $10^3$ GeV.
Higher-dimension operators might give extra contributions to the mass of this
axion field~\cite{bd1}.

One may easily extract from the scalar potential the masses of the
dilaton and of the moduli, which are in particular relevant for
cosmology. One finds for the moduli a mass
\be m_{t} \approx \left< {\pi
\rho_+\over6}{(b-b_+)\over(1+b\ell)}\right>.\label{eq:modmass} \ee
where $\rho_+$ is the hidden sector condensate with
the largest $\beta$-function coefficient $b_+$, and for the dilaton
\be m_d \sim {1 \over b_+^2} m_{\tilde G}. \ee 
In order to generate a hierarchy of order $m_{\tilde G}\sim
10^{-15}m_{Pl}\sim 10^3GeV$ we require~\cite{bgw2}
$b/b_+\approx 10$ in which case $m_t \approx 20 m_{\tilde G},\;m_d\sim 
10^3m_{\tilde G}$, which may be sufficient to solve the so-called
cosmological moduli problem.  

It is also straightforward to determine the soft supersymmetry 
breaking terms, that
are generated at the condensation scale $\mu_{cond} = \langle\rho_+^{1\over3}
\rangle$, in our model ~\cite{bgw2}. The gaugino masses are
\be m_{\lambda_b} = -\left<{g^2_b(\mu_{cond})\over8\ell^2}
\(\ell{\pp g\over\pp\ell}+1\)\sum_a \bar{u}_a\(1 + b_a\ell\)\right> 
\approx -{3\over8}{g^2_b(\mu_{cond}) b^2_+\over1 + b_+\langle\ell\rangle }
\left<\bar{u}_+\right>. \ee
The soft terms in the scalar potential are sensitive to the -- as yet 
unknown -- details of matter-dependent contributions to string 
threshold corrections and to
the Green-Schwarz term. We neglect the former,\footnote{If the threshold
corrections are determined by a holomorphic function, they cannot contribute to
scalar masses.} and write the Green-Schwarz term as
\be V_{GS} = b\sum_Ig^I + \sum_Ap_Ae^{\sum_Iq^A_Ig^I}|\Phi^A|^2 
+ O(|\Phi^A|^4), \ee
where the $\Phi^A$ are gauge nonsinglet chiral superfields, the $q^I_A$ are
their modular weights, and the full K\"ahler potential reads
\be K = k(V) + \sum_Ig^I + \sum_Ae^{\sum_Iq^A_Ig^I}|\Phi^A|^2 + O(|\Phi^A|^4).
\ee
With these assumptions the scalar masses and cubic ``A-terms'' are given,
respectively, by
\bea m_A^2 &=& {1\over16}\left<\left|\sum_a u_a{(p_A - b_a)\over
(1 + p_A\ell)}\right|^2\right> \approx {1\over16}\left<\rho_+^2{(p_A - b_+)^2
\over(1 + p_A\ell)^2}\right>, \nonumber \\ 
V_A(\phi) &=& { 1\over 4}e^{K/2}\sum_{a,A}\bar{u}_a\phi^AW_A(\phi)\[
{p_A-b_a\over1+p_A\ell}  + b_a-\(\ell{\pp g\over\pp\ell}+1\)
{1+b_a\ell\over3\ell} \] + {\rm h.c.}\nonumber \\
&\approx& { 1\over 4}e^{K/2}\bar{u}_+\[\sum_A{p_A-b_+\over1+p_A\ell}
\phi^AW_A(\phi) + {3b_+\over1+b_+\ell}W(\phi) \] + {\rm h.c.}, \eea
where $\phi = \Phi\lowest$ and $W(\Phi)$ is the cubic superpotential for chiral
matter. The scalar squared masses are positive and independent of their 
modular 
weights by virtue of the fact that $<F^I>$ vanishes in the vacuum. They
are universal -- and unwanted flavor-changing neutral currents are thereby 
suppressed -- if their couplings to the Green-Schwarz term are universal, in
which case the A-terms reduce to 
\be V_A(\phi) \approx {3\over4}e^{K/2}W(\phi)\bar{u}_+{p_A\(1+2b_+\ell\)-b^2_a
\ell\over(1+p_A\ell)(1+b_+\ell)} + {\rm h.c.} \equiv Ae^{K/2}W(\phi)+ 
{\rm h.c.}. \ee
If the Green-Schwarz term is independent of the matter fields $\Phi^A$,
$p_A = 0$ and we have $m_A = m_{\tilde G},\;A \approx 2m_{\lambda}$.  A
plausible alternative is that the Green-Schwarz term depends only on
the radii $R_I$ of the three compact tori that determine the untwisted sector
part of the K\"ahler potential (17): 
$$ K = k(V) - \sum_I\ln(2R_I^2) + O(|\Phi^A_{\rm twisted}|^2),$$ where
$2R^2_I = T^I + \bar{T}^I -
\sum_A|\Phi^A_I|^2$ in string units. In this case $p_A = b$ for the
untwisted chiral multiplets $\Phi^A_I$ and the 
untwisted scalars have masses comparable to the moduli masses: 
$m_A = m_t/2 \approx A/3$.\footnote{Scenarios in
which the sparticles of the first two generations have masses as high as
$20$ TeV have in fact been proposed~\cite{cohen}.} 
Finally, we note that if $b_+ \approx b/10 
\approx 1/30$, gaugino masses are suppressed relative to the gravitino mass at 
the condensation scale $\mu_{cond}\sim 10^{-4}m_{Pl}$: 
$m_\lambda \sim m_{scalar}/40$. If there is a sector with $p_A =
b$ and a Yukawa coupling of order one involving $SU(3)$ (anti-) triplets ({\it
e.g.} $\bar{D}DN$, where $N$ is a standard model singlet), its two-loop
contribution to gaugino masses~\cite{twoloop} can be more important than the 
standard one-loop
contribution, generating a physical mass for gluinos that is well within
experimental bounds for $m_{\tilde G} \sim$ TeV. Such a coupling could also
generate a $vev$ for $N$, thus breaking possible 
additional $U(1)$'s at a scale $\sim 10$ TeV. The phenomenologically
required $\mu$-term of the MSSM may also be 
generated by the $vev$ of a Standard
Model gauge singlet or by one of the other mechanisms that have been
proposed in the literature~\cite{muterm}. Clearly, a better
understanding of the $\Phi$-dependence of the string scale gauge coupling
functions is required to make precise predictions for soft supersymmetry
breaking.  Neverthless our model suggests soft supersymmetry breaking
patterns that may differ significantly from those generally assumed in
the context of the MSSM.  Phenomenological constraints such as current
limits on sparticle masses, gauge coupling unification and a charge and
color invariant vacuum can be used to restrict the allowed values of the
$p_A$ as well as the low energy spectrum of the string effective field theory. 

String non-perturbative corrections necessary to stabilize the dilaton
could make significant corrections to the unification of gauge
couplings.
The functions $f(\ell)$ and $g(\ell)$ introduced above and the threshold
corrections whose form is dictated by $T$-duality invariance contribute as
follows to the value of couplings at unification:
\bea
g_a^{-2} (\mu_s) &=& g_s^{-2} + {C_a \over 8 \pi^2} \ln (\lambda e)
-{1 \over 16 \pi^2} \sum_I b_a^I \ln (t^I + \bar t^I)|\eta^2(t^I)|^2,\\
g_s^{-2} &=& {f+1 \over 2 \ell}, \; \; \; \; \mu_s^2 = \lambda g_s^2
m_{Pl}^2,
\eea
with
\be
\lambda = {1\over 2} e^{g-1} (f+1)
\ee
Let us note however that this parameter is worth $1/(2e)\sim .18$ in the
perturbative case and $e^{-1.65} \sim .19$ in the one condensate model. 

We stress that the dependence on the radii moduli $T^I$ does not allow an
interpretation of the unification scale as the inverse radius of 
compactification. While the result (20) has been derived only for orbifold
compactifications, its large $T^I$ limit is consistent with the behavior found 
in the large $T^I$ limit of Calabi-Yau compactification. (Note that our moduli 
are fixed at the self-dual point, therefore far from this limit.)

To conclude, we would like to stress that the model presented above is
certainly not final and some of the results obtained, especially on the
low energy sector of the theory, may be modified. Possible sources of
modification are the presence of an anomalous $U(1)$ symmetry \cite{u1} or
a constant term in the superpotential that breaks the modular invariance
\cite{wr,horava}. Our model should be understood as  an existence proof 
of the fact that, even when nonperturbative effects play an
important role --especially in the stabilisation of the dilaton-- one
may compute in a reliable way  quantities of relevance in the low energy
world. This opens the way to a discussion of a certain number of issues
not addressed here, in particular the cosmology of the dilaton, the
axion and the moduli. Moreover, we have shown on our example that
predictions may be somewhat different from what appears to be the
standard lore. Finally, if one assumes that  the strongly
interacting string  is described along the lines of M-theory \cite{hw}, one 
may obtain information about non-perturbative contributions \cite{horava} 
but we expect that the general picture for low energy physics
will not be modified drastically.

\vskip .8cm
\noindent {\bf Acknowledgements}
\vskip .5cm
We thank Jim Gates, Petr Ho\v rova, Hitoshi Murayama and 
Hirosi Ooguri for discussions.
PB and MKG would like to acknowledge support from the Miller Institute
for Basic
Research in Science.  This work was supported in part by the Director, Office 
of Energy Research, Office of High Energy and Nuclear Physics, Division of 
High Energy Physics of the U.S. Department of Energy under Contract 
DE-AC03-76SF00098 and in part by the National Science Foundation under 
grant PHY-95-14797.

\end{document}